\documentstyle[11pt,paspconf,epsf]{article}

\begin{document}

\title{High Rydberg State Carbon Recombination Lines from
       Interstellar Clouds}
\author{K.R. Anantharamaiah$^{1,2}$ and Nimisha G. Kantharia$^1$}
\affil{$^1$Raman Research Institute, Bangalore 560 080, India}
\affil{$^2$National Radio Astronomy Observatory, Socorro, NM 87801, USA}
\begin{abstract}

We report observations of carbon recombination lines near 34.5 MHz
($n\sim 578$) and 325 MHz ($n\sim 272$) made towards Cas~A, the
Galactic centre and about ten other directions in the galactic plane.
Constraints on the physical conditions in the line forming regions are
derived from these and other existing observations. The CII regions
that produce the low-frequency lines are most likely associated with the
neutral HI component of the ISM.

\end{abstract}

\keywords{ISM:clouds, radiolines:ISM, line:profiles, Galaxy:general}

\section{Introduction}

Very low-frequency ($\nu < 100$ MHz) recombination lines of carbon
which arise from electronic transitions at high quantum number states
($n > 400$) are formed in a relatively cool (T$<$100 K) and partially
ionized widespread component of the ISM (Payne et al 1994, Erickson et
al 1995). These lines were first discovered by Konovalenko and Sodin
(1980) and identified by Blake et al (1980).  Below $\sim$150 MHz
($n>350$), the lines are seen in absorption and they turnover to
emission above $\sim$200 MHz (Payne et al 1989).  The width of the
lines increase dramatically towards lower frequencies which is caused
by pressure and radiation broadening.  The turnover frequency and
line-broadening are good indicators of physical conditions ($n_e$,
$T_e$ and $T_R$) in the line forming region. The occurrence of emission
lines from cold clouds against the `hot' radiation from Cas~A is
direct evidence for stimulated emission (Payne et al 1989). We report
here new observations of these lines near 34.5 MHz and 325 MHz towards
Cas A, the Galactic Centre and a number of other directions in the
galactic plane.

\section{Observations and Results}

Observations at 34.5 MHz were made using the 1.4 km $\times$ 20 m EW arm
of the T-shaped dipole array at Gauribidanur near Bangalore. The 530 m
$\times $ 30 m Ooty Radio Telescope was used for observations near 325
MHz. At 34.5 MHz, eight recombination line transitions ($n = 571-579$)
were observed simultaneously with a velocity resolution of $\sim 4.5$
km s$^{-1}$ and the spectra were averaged. The telescope beam was $21'
\times 25^{\circ}$. At 325 MHz, four transitions ($n = 270-272$) were
observed simultaneously with a velocity resolution of 1.8 km s$^{-1}$
and observations were made with two angular resolutions
$2^{\circ}\times 6'$ and $2^{\circ}\times 2^{\circ}$.

 \begin{figure}[t]
 \plotfiddle{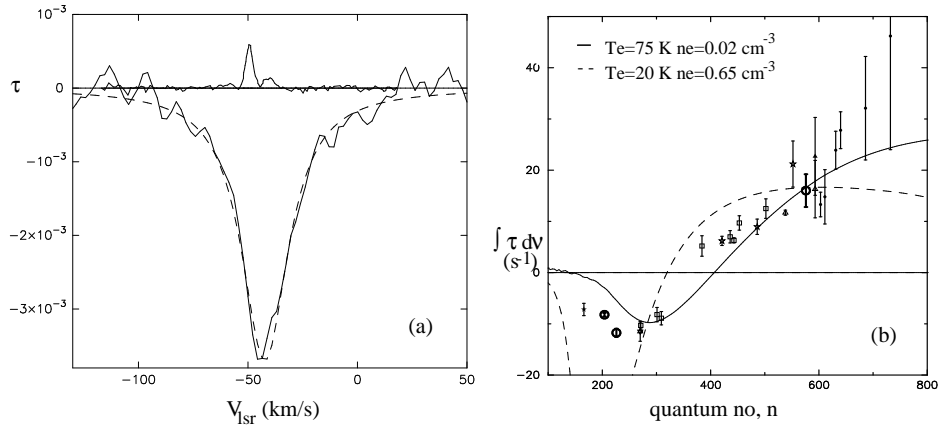}{2.0in}{0}{70}{70}{-200}{-330}
 \caption{(a) Carbon lines  towards Cas~A at 34.5 and 700 MHz.
 (b) Warm and Cold gas models with all available data points towards Cas~A}
 \end{figure}

Fig 1a shows the recombination line spectrum at 34.5 MHz towards Cas~A
with an effective integration time of $\sim$ 400 hrs. The spectrum
clearly shows the presence of broad wings expected from radiation and
pressure broadening. A Voigt profile fitted to the spectrum with a
Doppler width of 3.5 km s$^{-1}$ and a Lorentzian width of 25.4 km
s$^{-1}$ is superposed. Fig 1a also shows an emission spectrum
observed at 770 MHz using the 140 ft telescope in Green Bank
(Kantharia et al 1998). The differences in line shape and strength
between the two frequencies are striking. The high frequency line is
only Doppler broadened and shows two components of the Perseus arm in
emission.  The two components merge into a single feature at 34.5 MHz
due to line broadening.

Fig 2 shows spectra observed towards the galactic centre and a few
other positions in the galactic plane near 34.5 and 325 MHz. The lines
were detected towards 9 directions in the inner galaxy. One
significant difference in these spectra compared to Fig 1a
is that the line widths are similar at both frequencies. These line
widths are also similar to those observed at 75 MHz by Erickson et al
(1995).  Absence of line broadening at lower frequencies implies
that radiation and pressure broadening are not significant.  The line
widths are however large at both frequencies (20-30 km s$^{-1}$) which
is most likely due to differential galactic rotation.  The phenomenon
of turnover from absorption to emission is observed in all the cases.

 \begin{figure}[h]
 \plotfiddle{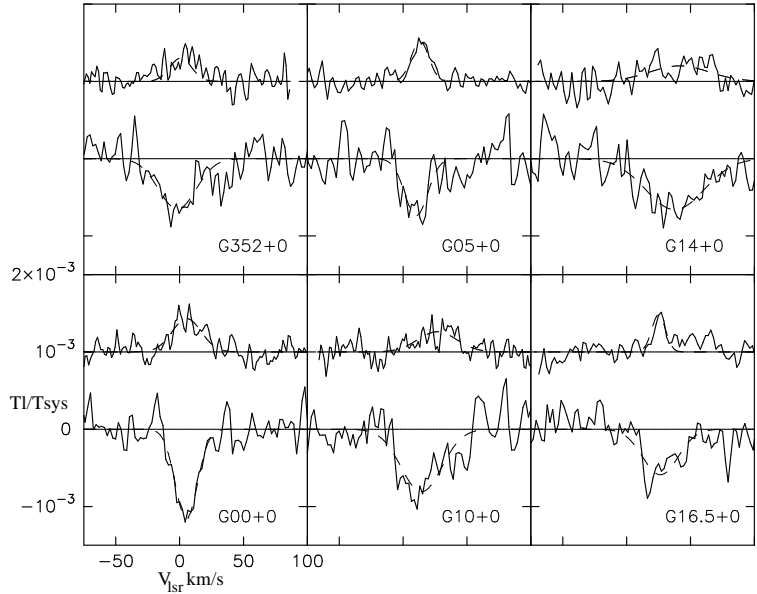}{2.8 in}{0}{60}{60}{-200}{-90}
 \caption{Carbon lines at 34.5 MHz and 325 MHz (upper spectrum)
 towards selected positions in the galactic plane.}  
 \end{figure}

\section{Constraints on the Physical Conditions}

If the entire Lorentzian width of 25.4 km s$^{-1}$ towards Cas~A is
attributed to radiation broadening, then we get an upper limit to the
radiation temperature at 100 MHz, $T_{R100} \le 3600$ K. Since the
brightness temperature of Cas~A at 100 MHz is $\sim 10^7$ K, the line
forming region cannot be too close to Cas~A. The implied lower limit
to the distance between Cas~A and the cloud is 115 pc. On the other
hand, if the cloud is farther away from Cas~A, then much of the
line width is due to pressure broadening. Two possible
combinations of electron temperature and density ($T_e, n_e$) that can
account for the line width are (75 K, 0.15 cm$^{-3}$) and (20 K, 0.65
cm$^{-3}$). The solutions are not unique. The observed line width thus
gives constraints on the combination of $T_e, n_e$ and $T_{R100}$.

Two types of models are generally considered for the line forming
regions (Konovalenko 1984, Sorochenko and Walmsley 1991, Payne et al
1994): the {\em warm gas models} with $T \sim 50 - 100$ K and
association with HI regions and {\em cold gas models} with $T < 20 $ K
and association with molecular gas. The models are evaluated by
computing the expected variation of integrated line strength with
quantum number and comparing with the observed values. The
combinations of $T_e, n_e$ and $T_R$ are chosen to be consistent with
the constraints from the line widths. The effects of a
dielectronic-like recombination on the population of high-$n$ states
of carbon are taken into account while calculating the expected line
strength (Walmsley and Watson 1982). Fig 1b shows all the available
data towards Cas~A gathered from the literature by Payne et al
(1994). The solid line shows the predictions of a warm gas model with
$T_e$ = 75 K, $n_e=0.02$ cm$^{-3}$, $T_{R100}$=3200 K and EM = 0.012
pc cm$^{-6}$. The dashed line is a cold gas model with $T_e$ = 20 K,
$n_e$ = 0.65 cm$^{-3}$, $T_{R100} = 800$ K and EM = 0.012 pc
cm$^{-6}$. The curves are normalized to the observed data point at
34.5 MHz. Clearly the cold gas model does not provide a good fit to
the data. The warm gas model provides fairly good fit to the
observations. However, the kinetic pressure implied by the parameters
is high - $nT\sim 10^4$ cm$^{-3}$ K, although it is still less than
the total hydrostatic pressure in the mid-plane.  The success of the
warm gas model suggests that the carbon line forming regions may be
associated with the cold neutral HI component of the ISM.  Kantharia
et al (1998) have presented a comparison of the spatial distribution
of C270$\alpha$, 21 cm HI and 3mm $^{12}$CO emission over the face of
Cas~A which supports this conclusion.

Deriving constraints for the lines detected in the inner Galaxy is
limited by (1) availability of few data points; typically one other
data point at 75 MHz (Erickson et al 1995) is available and (2) the
unknown angular size of the line region. The observations at 325 MHz
with two angular resolutions indicate that the line forming regions
are likely to be larger than $2^\circ$. In Fig 3, we show possible
warm gas models for three directions assuming the angular size to be
$2^\circ$. Model parameters are indicated in each frame.  The
path lengths through the gas implied by these models ranges from a few
tens of pc to more than a kpc and thus a distributed medium is
possible. The warm gas models were constrained to provide pressure
balance with the ISM. No warm gas model could be fitted for cloud
sizes larger than $\sim 4^\circ$. On the other hand if the cloud sizes
are in the range $5^\circ - 20^\circ$, then cold gas models can be
fitted to the data. However these models imply path lengths of $\sim
0.05$ pc which would be in conflict with the observed large line
widths which are most likely due to galactic rotation. The kinetic
pressure in the cold gas models are typically a factor of ten larger
than the average ISM value. The overall evidence seem to favour the
warm gas models.

 \begin{figure}[t]
 \plotfiddle{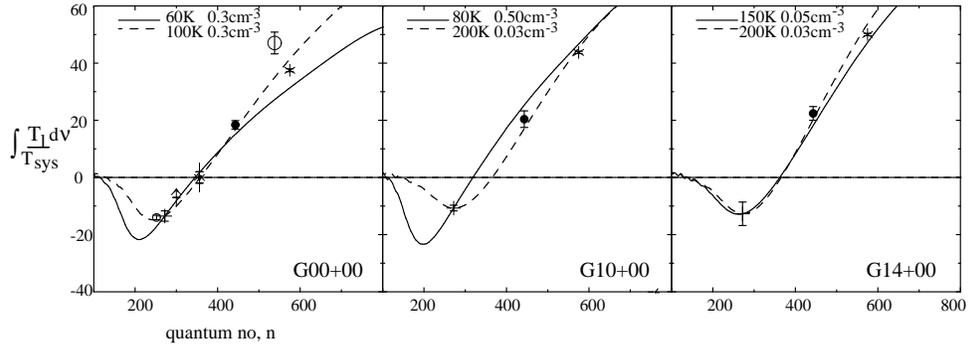}{1.5 in}{0}{80}{80}{-240}{-360}
 \caption{Warm gas models for three positions in the galactic plane.}
 \end{figure}

\end{document}